\documentclass[rnote]{aa} 
\usepackage{graphicx}
\usepackage[varg]{txfonts}
\usepackage{longtable}

\usepackage{natbib}

\def\ltsima{$\; \buildrel < \over \sim \;$}
\def\simlt{\lower.5ex\hbox{\ltsima}}
\def\gtsima{$\; \buildrel > \over \sim \;$}
\def\simgt{\lower.5ex\hbox{\gtsima}}
\begin{document}
   \title{The surface brightness profile of the remote cluster NGC~2419}

   \author{M. Bellazzini\inst{1}}

   \offprints{M. Bellazzini}

   \institute{
             INAF - Osservatorio Astronomico di Bologna,
              Via Ranzani 1, 40127 Bologna, Italy\\
          \email{michele.bellazzini@oabo.inaf.it}             }

     \authorrunning{M. Bellazzini}
   \titlerunning{The surface brightness profile of NGC~2419}

   \date{Accepted by A\&A, July 17, 2007}

\abstract
{It is well known that the bright and remote Galactic globular cluster 
NGC~2419 has a very peculiar structure. In particular its half-light radius is
significantly larger than that of ordinary globular clusters of similar
luminosity, being as large as that of the brightest nuclei of dwarf elliptical
galaxies.}
{In this context it is particularly worth to check the
reliability of the existing surface brightness profiles for this cluster 
and of the available estimates of its structural parameters.}
{Combining different datasets I derive the surface brightness profile going 
from the cluster center out to $\simeq 480\arcsec$, i.e. $\simeq 25$ core radii 
($r_c$).
The profile of the innermost $21\arcsec$ has been obtained from aperture
photometry from four different Hubble Space Telescope
ACS/WFC images. Outside of this radius, the profile has been obtained from 
star counts.}
{The newly obtained surface brightness profile is in excellent agreement with
that provided by Trager, King \& Djorgovski for $r\ga 4\arcsec$.
The new profile is best fitted by a King model having $r_c=0.32\arcmin$
($\sim$ 5\% smaller than previous estimates),
central surface brightness $\mu_V(0)=19.55$ and concentration $C=1.35$.
Also new independent estimates of the total integrated V magnitude 
($V_t=10.47 \pm 0.07$) and of the half-light radius 
($r_h=0.96\arcmin \pm 0.2\arcmin$) have been obtained.
 The average ellipticity in the range $5\arcsec \le r\le 120\arcsec$ is 
$\langle \epsilon \rangle =0.19\pm 0.14$. 
 If the four points of the ellipticity profile that
deviates more than $2\sigma$ from the overall mean are excluded, 
$\langle \epsilon \rangle =0.14\pm 0.07$ is obtained.
}
{The structure of NGC2419 is now reliably constrained by 
(at least) two fully independent observational profiles that are in 
good agreement one with the other. Also the overall agreement between 
structural parameters independently obtained by different authors 
is quite satisfying.}


   \maketitle
%

\section{Introduction}

NGC~2419 is one of the brightest globular clusters (GC) of the Milky Way and it
is located at very large distance from the center of the Galaxy ($R_{GC}=91.5 $
kpc, see Harris \cite{harris}). Moreover, it is well known that it has a
quite peculiar structure: its half-light radius is much larger 
(by a factor of $\sim 5$)
than that of other GCs of the same luminosity, being as large as that of the
largest nuclei of dwarf elliptical galaxies,  or the scale sizes of Ultra
Compact Dwarfs (Mackey \& van den Bergh
\cite{macksyd}, Federici et al. \cite{F07}, 
 Evstigneeva et al. \cite{evsti}
and references therein; see Fig.~13 of van den Bergh \cite{vdb} for a direct 
visual demonstration of the extra-large size of the cluster). 
For these reasons, it has been proposed that NGC~2419 is in
fact the remnant nucleus of a dwarf elliptical satellite of the Milky Way that
was partially disrupted by the Galactic tidal field (Mackey \& van den Bergh
\cite{macksyd}, but see Ripepi et al. \cite{ripepi} for arguments against this
hypothesis). 
The structure of GCs is usually described with the parameters of
the King \cite{king} model that best fits their Surface Brightness (SB) profile,
that are the central SB $\mu(0)$, the {\em core radius} $r_c$ and
the concentration $C=$log$(r_t/r_c)$, where $r_t$ is the {\em tidal radius}
(see King 1966, hereafter K66). Other fundamental parameters, not necessarily
linked to K66 models, are the total absolute magnitude, 
usually reported in the V
passband, $M_V$, and the half-light radius $r_h$. 

Independently of the actual origin of NGC~2419, its peculiar structure is
clearly an interesting subject of study. In this sense, it seems particularly
important to check the reliability of the parameters that makes this clusters so
special and worth of further investigations, that is its structural parameters
and the SB profile they are obtained from. There are
several independent estimates of (at least some) structural parameters of
NGC~2419 (see, for example, Peterson \& King \cite{pk75}, Natali, Pedichini \&
Righini \cite{natali}, hereafter NPR, 
Cohen et al. \cite{coh}, and references therein), but the
only publicily available SB profile is that provided by Trager, King \&
Djorgovski (1995, hereafter TKD) by assembling the observational material
previously obtained with various methods and by various authors. The same
observed profile was later re-analyzed by McLaughlin \& van der Marel (2005,
hereafter MvdM). 
While there is no particular reason to doubt of the reliability of the TKD
profile, a check with a completely independent observed profile would be 
clearly valuable. 
I have obtained such a profile by combining three different datasets
and two different techniques. This short note is aimed at reporting on this
newly obtained profile and structural parameters and on the comparison with
previously available data. 

\section{Data Analysis}

The first dataset adopted for the analysis is a set of sky-subtracted drizzled
images taken with the Advanced Camera for Surveys / Wide Field Channel 
(ACS/WFC) camera on board of the Hubble Space Telescope (HST): 
a F475W image with $t_{exp}=680$ s, a F606W image with $t_{exp}=676$ s, 
a F814W image with $t_{exp}={676}$ s, and a F850LP image 
with $t_ {exp}=1020$ s\footnote{These are {\em total} exposure times of the {\em
combined} drizzled images. The inspection of the images reveal that there is no
heavily saturated star in the innermost $20\arcsec$ region where aperture
photometry has been performed.},
that were retrieved from the HST archive
(images j8io01031\_drz, j8io01081\_drz, j8io01071\_drz, and, j8io01041\_drz, 
respectively, from the GO9666 program, P.I. L. Gilliland).
The region within $21\arcsec$ from the cluster center is fully enclosed in all
these images. I obtained the SB profile in this region by aperture photometry on
concentric, $3\arcsec$ wide, annuli in all the four images\footnote{Using the
XVISTA package, see {\tt http://astronomy.nmsu.edu/holtz/xvista/index.html} and
Lauer \cite{lauer}.}. By definition, a surface density profile obtained with
aperture photometry is completely unaffected  by incompleteness, as opposite to
profiles obtained from star counts. Therefore this portion of the profile will
provide the fundamental benchmark to check the effects of radial variations of 
the completeness in the innermost part of the profile derived from star counts
(see below). Moreover, the surface photometry can be reported  
to an absolute scale by means of the photometric Zero Points (ZP) of 
Sirianni et al. \cite{sirianni}. Normalizing the profile from star counts to the
inner profile obtained from aperture photometry, the whole composite profile can
be reported to the same absolute photometric scale.

I searched the center of symmetry of the cluster by computing the light density
over apertures of radius $=4\arcsec$ in many different positions around the
apparent center of the cluster; the maximum was found at the image (pixel) 
coordinates ($x_0$,$y_0$)=(610,2895). 
Taking the astrometric solution embedded in the
images as a reference, this is fully consistent with the position of the 
center that is available in the literature (Harris \cite{harris}). Since the
adopted center of the raster of annuli may be critical for the aperture
photometry, I derived the profile by adopting seven different positions of the
center of the annuli, i.e. (x,y)=($x_0\pm 10$ px,$y_0$), 
(x,y)=($x_0$,$y_0\pm 10$ px), (x,y)=($x_0- 10$ px,$y_0+ 10$ px),
(x,y)=($x_0+ 10$ px,$y_0- 10$ px), and, obviously (x,y)=($x_0$,$y_0$).
The final SB value in each annulus is the average of the seven values obtained
with the different assumptions on the coordinates of the center, and their
standard deviation is the adopted uncertainty. The F606W and F814W profiles have
been calibrated in the VEGAMAG system, while the F475W and F850LP 
profiles have been
calibrated in the ABMAG system, using the ZPs by Sirianni \cite{sirianni}.
The reason of the adoption of different systems will become clear below. 

\begin{figure}
   \centering
   \includegraphics[width=9cm]{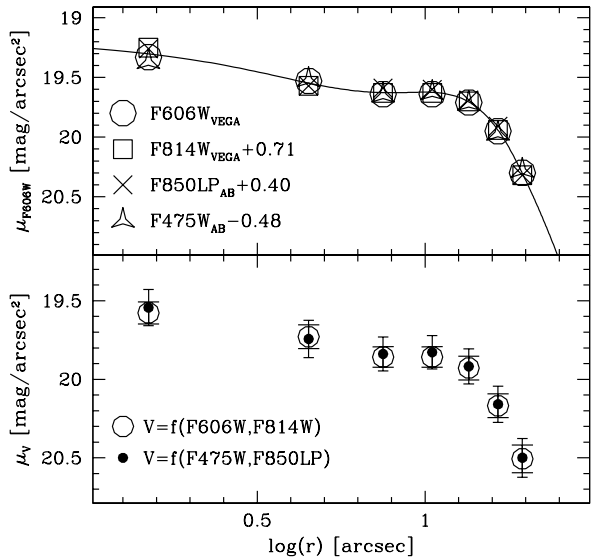}
    \caption{Upper panel: observed Surface Brightness profile from aperture
    photometry within $r\le 21\arcsec$ of the four ACS images considered
    here. An arbitrary normalization constant have been added to the F814W,
    F850LP and F475W profiles to match the zero point of the F606W profile.
    Note the virtually perfect match among the {\em shapes} of the various
    profiles.
    The continuous line is a cubic spline interpolated on the average of the
    four SB values.
    Lower panel: derived SB profiles in V obtained from (a) the F606W and F814W
    profiles (VEGAMAG system) {\em and} the transformation by G06 
    (large open circles), and (b) the F475W and F850LP profiles 
    (ABMAG system) {\em and} the transformation by F07 (filled circles). The
    error bars include uncertainties in the zero points and in the
    transformations. Note the {\em excellent} match between the two {\em fully
    independent} profiles.
    }
           \label{bsurf}
    \end{figure}

The aperture photometry profiles in the various passbands are compared in the 
upper panel of Fig.~\ref{bsurf}; arbitrary constants have been added to the 
F814W, F475W and F850LP profiles to match the Zero Point of the F606W profile. 
In this way
the {\em shape} of the different profiles can be properly compared. The agreement
among the different profiles is excellent. All previous determinations of the
profile of NGC2419 were obtained from ground based data, in many cases with
seeing width significantly larger than $1\arcsec$. The profile for $r\le
21\arcsec$ obtained here from ACS data is clearly superior to all of them and
should be taken as the reference, in this range of distances from the cluster
center. For each profile we
measure the Half Width at Half Maximum (HWHM) length, that is a good initial
proxy for $r_c$. From the average of the four values we obtain HWHM$=18.0\arcsec
\pm 1.0\arcsec$.

Galleti et al. (2006, hereafter G06) 
obtained a relation to transform F606W magnitudes into standard V
magnitudes\footnote{In the Johnson-Kron-Cousins system, based on Landolt
\cite{land} standard stars.}, as a function of F606W-F814W color 
(VEGAMAG system). The V profile
obtained by applying this relation to the observed F606W, F814W profiles is
plotted in the lower panel of Fig.~\ref{bsurf} (large empty circles).
On the other hand, Federici et al. (2007, hereafter F07) obtained a relation to 
transform F475W 
magnitudes into standard V, s a function of F475W-F850LP color (ABMAG system).
The V profile obtained by applying this relation to the observed F475W, F850LP 
profiles is also plotted in the same panel of Fig.~\ref{bsurf} 
(filled circles). In both cases, uncertainties in the ZP and 
in the transforming
relations have been included in the error bars. {\em The agreement between the
two fully independent V profiles is excellent.} This further check, strongly
support the reliability of the absolute surface photometry in the innermost
$21\arcsec$ of NGC~2419. In the following I will take the V profile obtained 
from the F606W,F814W profiles as the reference, since it is based on a
less-scattered, more reliable photometric transformation (G06). 
From the derived V profile, $\mu_V(0)=19.55\pm 0.10$ is obtained, $\sim 0.2$ mag
brighter than reported by TKD, but $\sim 0.1$ mag fainter than what found by
MvdM from the same data as TKD. For the reasons outlined above, also the value
of $\mu_V(0)$ obtained here should be regarded as very reliable.

Photometry of individual stars in the ACS images considered here 
were obtained by G06 and F07, to
derive the relations between photometric systems. Here I use their 
F475W,F850LP Color Magnitude Diagram (CMD) to select the stars to count to
extend the SB profile outside the small region covered by integrated aperture
photometry. In the following we will refer to the adopted catalogue of positions
and photometry of individual stars as to the ACS sample. 

\subsection{Ellipticity from integrated photometry}

The ACS images are too small and off-centered to be used to obtain a global
estimate of the ellipticity of the cluster. I have retrieved from the 
Italian Center of Astronomical Archives\footnote{IA2, at 
{\tt http://wwwas.oats.inaf.it/IA2/}} a t$_{exp}$=240 s V image of NGC2419 taken
with the DOLORES camera\footnote{See 
{\tt http://www.tng.iac.es/instruments/lrs/}} at the Telescopio Nazionale Galileo (TNG), at the
observatory Roque de Los Muchachos, in La Palma, Canary islands (Spain). The
image was acquired in 2004, January 27, during a test session; the seeing was
$\sim 1\arcsec$ FWHM. The pixel scale is $0.275\arcsec$px$^-1$.
The image was corrected for bias and flat field with
standard IRAF procedures and it was used to study the ellipticity of the
cluster over the  range $5\arcsec \le r \le 120\arcsec$, i.e. 
approximately out to two times the half-light radius of the cluster 
(see below).

The ellipticity ($\epsilon=1-{b\over{a}}$, where $a$ and $b$ are the 
semi-major and
semi-minor axes, respectively) was computed by finding the ellipses that best
fits the light distribution, while keeping fixed the center of the ellipses, 
using the XVISTA task {\em profile}, as described in detail in F07. 
This task computes the (elliptical) light 
profile adopting one pixel step: to reduce the noise I averaged all the 
derived quantities over $\pm 5\arcsec$ wide bins, as done in F07. 
The shape of the derived SB profile was in good agreement with the ACS profile,
in the region of overlap.

\begin{figure}
   \centering
   \includegraphics[width=9cm]{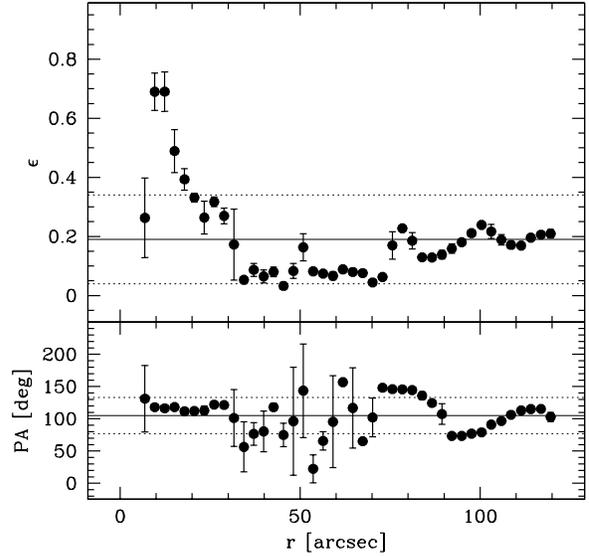}
    \caption{Upper panel: ellipticity as a function of distance from the cluster
    center. The thin continuous line marks the value of the mean ellipticity, 
    averaged over the whole radial range, the dotted lines are at $\pm 1
    \sigma$ from the mean. Lower panel: Position Angle as a function of 
    distance from the cluster center. The lines have the same meaning as in the
    upper panel.}
           \label{elli}
    \end{figure}

The average ellipticity of NGC2419 over the considered radial range is
$\epsilon=0.19 \pm 0.15$, where the uncertainty is the standard deviation of the
distribution. The average Position Angle, computed from North (PA=$0\degr$)
toward East (PA=$+90\degr$), is $\langle PA\rangle=+105\degr \pm 28\degr$.
 The ellipticity and PA profiles are shown in Fig.~\ref{elli}. 
Except for a narrow peak in the range $10\arcsec \la r\la 20\arcsec$, the
ellipticity profile is always within $\pm 1\sigma$ from the overall mean.
A $2\sigma$ clipping average excludes the four points of the profile with the highest
ellipticity, leading to a slightly lower mean $\epsilon=0.14 \pm 0.07$.

\subsection{Star Counts}

\begin{figure*}
   \centering
   \includegraphics[width=16cm]{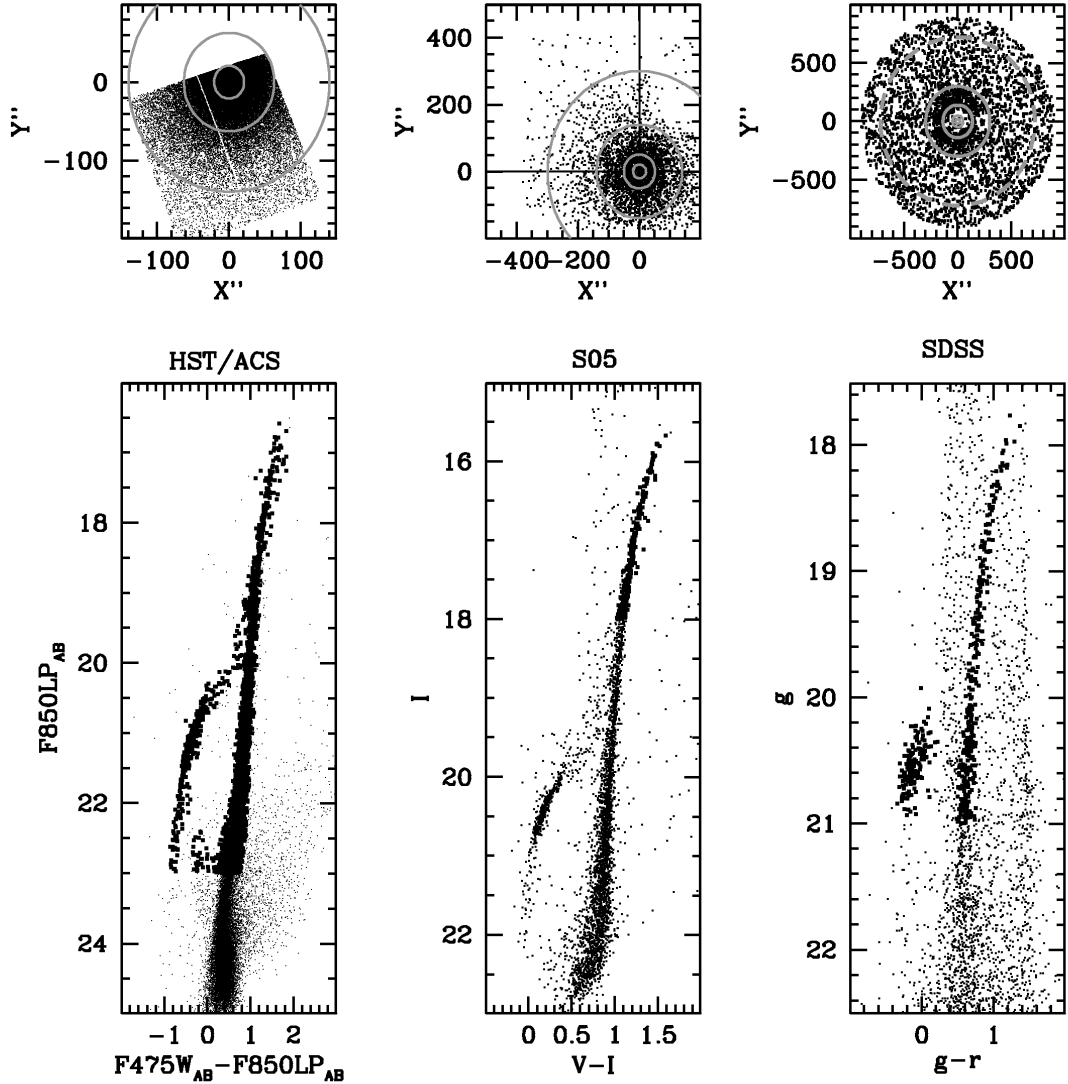}
    \caption{Samples used for the star counts (upper panels: sky-projected maps;
    lower panels Color Magnitude Diagrams). The stars selected for the analysis
    are plotted as heavier dots in the CMDs. The grey circles plotted in the
    maps marks the boundaries between regions covered by a given dataset and/or
    the regions of overlap among the various
    datasets that have been used to derive the SB profile. 
    In order of increasing radius:
    $r\le 21\arcsec$: Aperture Photometry on ACS images; 
    $21\arcsec<r\le 51\arcsec$: star counts on ACS data; 
    $51\arcsec<r\le 140\arcsec$: region of overlap between the ACS and the S05
    data (star counts); $140\arcsec <r\le 300\arcsec$: region of overlap
    between the S05 data (upper left quadrant) and the SDSS data (star counts);
    $r=720\arcsec=12\arcmin$: (dashed circle) outer boundary for the star counts
    from the SDSS data. The background level was estimated in the area outside 
    of this circle, up to $r=15\arcmin$.}
           \label{samp}
    \end{figure*}

In Fig.~\ref{samp} the photometric samples that have been used to derive the
surface density profile from star counts on concentric annuli are presented.
The innermost sample is constituted by the ACS/WFC photometry already described
above (left panels). An intermediate sample is provided by the accurate photometry
by Saha et al. (2005, hereafter S05; central panels); in particular this 
sample cover one full quadrant over a radial range that joins the ranges of the 
ACS sample and the outermost sample, that was retrieved from the Sloan Digital
Sky Survey (Adelman-McCharty et al. 2005, hereafter SDSS).     
Cluster stars are selected on the Color Magnitude Diagram (CMD) 
as shown in the lower
panels of Fig.~\ref{samp}. The circles overplotted on the upper-panels maps of
Fig.~\ref{samp} shows that there are generous overlapping 
regions between the adopted samples.

Surface density profiles were obtained from each sample, using different bin
widths, according to the different density of tracer stars of the different
samples and in each different radial range. 
Each sample is intended to cover a given radial range, and the
corresponding profiles are compared (and matched) 
in the regions of overlap (see
the circles superposed to the maps in the upper panels of Fig.~\ref{samp}), 
with
a bootstrap process that uses as a reference the innermost profile from aperture
photometry, and then, in turn, the ACS, the S05 and the SDSS samples.
The effects of radial
variations of the incompleteness can be checked by comparing 
with a profile that has been proven to be
unaffected by this problem: if the profiles have the same shape in the region
of overlap, the outer profile does not suffer from radially varying
incompleteness (see Federici et al. 2007). 
For example, in the case of the ACS sample, surface density profiles were
obtained from star counts by using different magnitude thresholds
($F850LP<20$, $F850LP<21$, ..., $F850LP<23$) to select the tracer stars
(brighter thresholds would correspond to more complete samples).
The star count profile from the $F850LP<20$ sample nicely matched the aperture
photometry profile in the innermost $10\arcsec$, hence it is free from
incompleteness problems in that range. The $F850LP<22$ profile, 
that suffered from significant incompleteness for $r< 8\arcsec$, matched 
the $F850LP<20$ profile for $r> 10\arcsec$, hence it was adopted in that range. 
Analogously, it was possible to adopt the $F850LP<23$ 
for $r\ge 27\arcsec$, since there was excellent match with the $F850LP<22$ 
profile for $r\ga 20\arcsec$. In this way all the various profiles, from the
various samples, were assembled into a single ``star counts'' profile that is
unaffected by radial variations in the completeness.

The contribution of the background to the surface density was subtracted from 
whole profile obtained from star counts; the background density was
estimated in the $720\arcsec \le r \le 900\arcsec$ region of the SDSS sample (see
the upper right panel of Fig.~\ref{samp}), well beyond the tidal radius of the
cluster ($r_t=522\arcsec$, TKD). 
Finally the profile was matched to
the V profile obtained from the aperture photometry to convert in calibrated SB
units, and, finally, the combined profile reported in Tab.~\ref{tab1} was 
produced (Tab.~\ref{tab1} is available only in electronic form, 
in the on-line edition of this note). The profile from integrated photometry
obtained from the TNG data (Sect.~2.1, above) was used to further check that the
overall combined profile was well-behaved in the region where aperture
photometry and star counts data have been joined. 

 The bright magnitude threshold adopted for the S05 and - to a lesser extent - 
for the SDSS sample (see
Fig.~\ref{samp}) implies that in the outer low-density 
regions of the cluster covered by these samples the uncertainty on the
derived SB can be quite large, in some case.
The accurate tracing of these parts of the profile would require much deeper 
wide-field photometry than what publicily available (see, for example, Ripepi et al.
2007).
Moreover I excluded from the final profile all the annuli where the number of
selected cluster stars were $<10$, hence the SB estimates of Tab.~\ref{tab1} 
are not necessarily regularly spaced.
Nevertheless, (a) the reported SBs for
$r\le 51\arcsec$ are, by far, the most accurate estimates presently available,
(b)  the new profile covers a much larger radial range with respect 
to previously available ones (TKD), i.e. 
$r\simeq 8\arcmin \simeq 1 r_t \simeq 25 r_c$, and, (c) it is obtained from
datasets never used before for this purpose, thus providing observational
constraints on the structure of NGC2419 that are fully independent from what 
was already available in the literature.

\section{Structural parameters for NGC~2419}

\begin{figure}
   \centering
   \includegraphics[width=9cm]{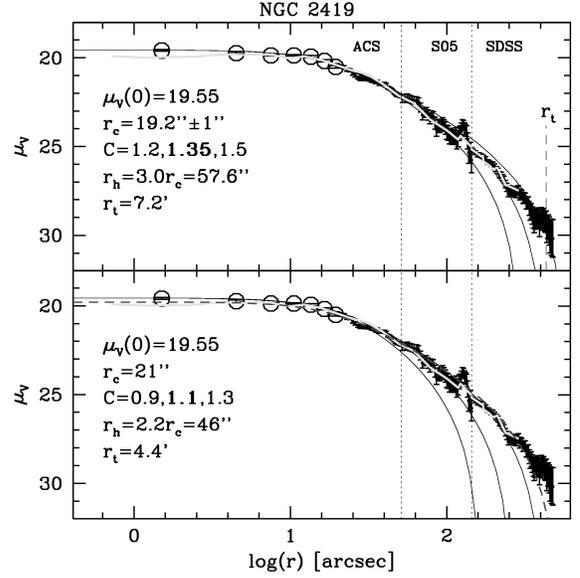}
    \caption{Comparisons between the SB profile derived in the present analysis
    (open circles: from aperture photometry; small dots with errorbars: from
    star counts), the composite profile adopted by TKD (light grey line), 
    and various King's models (thin continuous lines). 
    Upper panel: adopting the new estimates of $\mu_V(0)$ and $r_c$, 
    the observed profile is compared with King's models with C=1.2, 1.35, and 
    1.5, from left to right. The best-fitting C value is written in boldface.
    The vertical dotted lines marks the boundaries between the portions of the
    profile obtained from different datasets.  The dashed segment marks the
    position of the tidal radius of the best fit King model.
    Lower panel: the same as above but adopting the $r_c$ value of TKD. In this
    case there is no model providing a good fit over the whole radial range and
    a C as low as 1.1 is required to fit the profile out to $r\sim 100\arcsec$.
    The grey dashed line is a C=1.4 king's model, adopting $\mu_V(0)=19.77$, as
    done by TKD.}
           \label{fatti}
    \end{figure}

In Fig.~\ref{fatti} we compare the profile of Tab.~\ref{tab1}
with the theoretical profiles of isotropic single-mass K66 models,
and with the combination of observed profiles adopted by TKD.
In the upper panel, we adopt the central surface brightness and the HWHM values 
obtained in Sect.~2. The overall profile is best-fitted by a model with C=1.35,
in good agreement with the results by TKD and MvdM. Adopting this model, the
resulting core radius is $r_c=1.07HWHM=19.2\arcsec$, and the half-mass radius
is $r_{hm}\simeq 3.0r_c=57.6\arcsec$. Since the relaxation time of NGC2419 is
much larger than one Hubble time (Djorgovski \cite{djor}, MvdM), 
the effects of mass segregation should
be very small in this cluster, hence the half-mass radius should be a good proxy
for the half-light radius; for this reason I assume $r_h=r_{hm}=57.6\arcsec$.

It is very interesting to note that {\em the newly (and independently) derived
profile is in excellent agreement with the TKD profile for $r\ge4\arcsec$}, 
with, perhaps, the exception of a small wiggle at log(r)=1.4 in the region 
covered by ACS star counts. Given the initial purpose of the exercise reported
in this note, i.e. to check the reliability of existing profiles, this result
can be regarded as very reassuring. 
The only significant difference between the two profiles occurs at the 
innermost point:
the TKD profile shows a curious drop in the inner few arcseconds. This lead TKD
to derive a central SB much fainter than what found here, $\mu_V(0)=19.77$
instead of  $\mu_V(0)=19.55$. 
 Beyond the latest point of the TKD profile, for $r\ga 290\arcsec$, the observed 
profile show a slight change of slope and a SB excess with respect to the best-fit King 
model. The feature is suggestive of the possible presence of extra-tidal stars (see
Leon et al. \cite{leon}, F07, and also Ripepi et al. \cite{ripepi}). However, given the
typical uncertainties associated with this outer part of the profile, the feature is not
further discussed in the following and it is not considered in the comparisons
with TKD and with theoretical models.   

The dashed line in the lower panel of
Fig.~\ref{fatti} is a C=1.40 King's model with  $\mu_V(0)=19.77$: its comparison
with the observed profiles shows that
if such a faint central SB is adopted, a core radius as 
large as $r_c=21\arcsec$ is needed to obtain a good fit at large radii. 
Therefore, the difference
in the central SB is at the origin of the difference in the estimate of 
the core radius between the present analysis and TKD, that occurs  
in spite of  the good agreement between the two profiles. On the other hand, if
a core radius as large as $r_c=21\arcsec$ and $\mu_V(0)=19.55$ are
simultaneously adopted, it is no more possible to
obtain a good fit of the observed profile over the whole considered
radial range ($r\le 300\arcsec$) with a single King's model. A C=1.1 
model is preferable for $r\la 100\arcsec$, while a C=1.3 model is required 
at larger radii (Fig.~\ref{fatti}, lower panel).

Since there is no reason to expect a central drop in the surface brightness, 
given the reliability of the aperture photometry profile derived here from ACS
images, and since a unique fitting model for the whole radial range is
preferable to two, I finally adopt the solution presented in the upper panel of
Fig.~\ref{fatti}, and the associated structural parameters, 
reported in the first column of Tab.~\ref{tab2}. 

To compute the total apparent integrated V magnitude I first derived the
integrated magnitude within the central $20\arcsec$, through aperture
photometry on the F606W image, finding $V_{r\le20\arcsec}=12.27 \pm 0.05$, 
where the uncertainty is dominated by the uncertainties in the photometric 
zero points and  in the transformations from F606W to V. 
Then I integrated the K66 model that best fits the observed
profile (C=1.35) to estimate what fraction of the light is included in the
innermost $20\arcsec$, $F_{20}=L_{r\le20\arcsec}/L_{r\le r_t}=0.191$.
The total integrated magnitude is:

\begin{equation}
V_t = V_{r\le20\arcsec} +2.5log(F_{20}) = 10.47 \pm 0.07
\end{equation}

Adopting $(m-M)_0=19.60\pm 0.05$ and $E(B-V)=0.08$, from Ripepi et al.
\cite{ripepi}, the absolute integrated V magnitude is $M_V=-9.4\pm 0.1$.

\section{Summary}

I have derived a new surface brightness profile for the remote globular cluster
NGC2419 by using different datasets that were never used before for this
purpose. This allowed me to verify the reliability and the accuracy of both {\em
the shape} and {\em the photometric zero point} of previously existing profiles,
as well as the accuracy of available estimates 
of the structural parameters for this peculiar and very interesting
cluster. The main results of the present analysis can be summarized as follows:

\begin{itemize}
\item{} The agreement between the newly derived profile and the previously
available one by TKD is more than satisfying. The overall agreement among
the estimates of the main structural parameters from various sources is also
sufficiently good (see Tab.~\ref{tab2}). Therefore, it must be concluded that 
the structural parameters of NGC2419 are known with a good degree 
of reliability. 

\item{} The newly derived profile suggests a slight downward revision of the core
radius, with respect to previous estimates, i.e. 
$r_c=19.2\arcsec \pm 1\arcsec$ instead of $r_c\ga 20.5\arcsec$. 
The central SB derived here ($\mu_V(0)=19.55$) is more similar to the estimate 
by MvdM  ($\mu_V(0)=19.44$) than to that by TKD ($\mu_V(0)=19.77$).
The less well constrained structural parameter is the tidal radius, that is
extrapolated from the inner profile ($r\la 290\arcsec$).
The $r_t$ value found here is
slightly smaller with respect to previous analyses.
This may have some effect on some related issues, as the possible detection of
extra-tidal stars by Ripepi et al. \cite{ripepi}, and the estimate of the mass
of the Milky Way obtained in  Bellazzini \cite{mb}. 
The profile shows a slight break and a SB excess for $r\ga 290\arcsec$ that 
may be indicative of the presence of  an extra-tidal component 
(Leon et al. \cite{leon}).
The evidence is not conclusive and deserves to be followed-up with deeper
wide-field photometry.

\item{} The overall cluster shape is not so round, at least within
$r\simeq 120\arcsec$, a radius enclosing  significantly more than one 
half of the  whole cluster light.  
The average ellipticity in this range is $\epsilon =0.19 \pm 0.15$ 
 (or
$\epsilon =0.14 \pm 0.07$, if a $2\sigma$ clipping algorithm is applied), a
relatively high value for a globular cluster (see Barmby et al. 2007).
Quite elongated shapes seem to be a
typical carachteristic of bright and large-sized clusters 
(see F07 and Galleti et al. 2007).
The much lower values ($\epsilon = 0.03/0.04$, 
averaged over the whole cluster extent) found by Frank \& Fall \cite{FF} and 
White \& Shawl \cite{WS} may suggest a decrease of ellipticity at 
large radii.
 
\end{itemize} 
%
%
%
%
%

%
%
%
\begin{table}
\setcounter{table}{1}
\begin{minipage}[t]{\columnwidth}
\caption{Structural Parameters of NGC~2419}
\label{tab2}
\centering
\renewcommand{\footnoterule}{}  
\begin{tabular}{cccccc}
\hline \hline
   Parameter& this work & TKD& MvdM & NPR &units \\
\hline
HWHM         &$18.0\pm 1.0$ & --  &      & 19.2 &arcsec    \\
r$_c$        &$0.32\pm 0.02$ & 0.35 & 0.34    & 0.34 &arcmin    \\
C           & 1.35          & 1.4      &  1.38 & 1.5 & --       \\
r$_h$         &$0.96\pm 0.2  $ & 0.76     & 0.86 & &arcmin    \\
r$_t$         &$7.1 \pm 1.0 $ & 8.7      & 8.2  & 10.71\footnote{From Peterson \& King (1975).}&arcmin    \\
$\mu_V(0)$  &$19.55\pm 0.10$&19.77   & 19.44    &    &mag/arcsec$^2$\\
$V_t$       &$10.47 \pm 0.07$&10.26\footnote{From Peterson (1993).}     & 10.29& &mag        \\
$M_V$       &$-9.4\pm 0.1 $&-9.58     & & &mag        \\
$\langle\epsilon\rangle$       &$0.19\pm 0.15 $&    & & &    --    \\
$\langle\epsilon\rangle_{2\sigma}$       &$0.14\pm 0.07 $\footnote{Average obtained after the exclusion of the points of the ellipticity profile
that deviated more than  $2-\sigma$ from the overall mean.}&    & & &    --    \\
\hline
\end{tabular}
\end{minipage}
\end{table}

\begin{acknowledgements}
The financial support to this research by the INAF-PRIN05 through the 
Grant CRA 1.06.08.02 is acknowledged.
Based on observations made with the NASA/ESA Hubble Space Telescope, obtained 
from the Data Archive at the Space Telescope Science Institute, which is 
operated by the Association of Universities for Research in Astronomy, Inc., 
under NASA contract NAS 5-26555. These observations are associated with 
program GO9666.
Based on observations made with the Italian Telescopio Nazionale 
Galileo (TNG) operated on the island of La Palma by the Fundaci\'on Galileo 
Galilei of the INAF (Istituto Nazionale di Astrofisica) at the Spanish 
Observatorio del Roque de los Muchachos of the Instituto de Astrofisica de 
Canarias.
This research made use of data from the Sloan Digital Sky Survey. 
I'm grateful to Luciana Federici for an introduction to the basics of XVISTA.
This research made use of the NASA/ADS database.
\end{acknowledgements}

\bibliographystyle{aa}

\onltab{1}{
\onecolumn
\clearpage
\begin{scriptsize}

\begin{longtable}{l c c c c }
\caption{Observed surface brightness profile of NGC~2419} \\
\label{tab1}\\
\hline\hline 
   r   & $\mu_V$        & $err_{\mu_V}$    & Technique & Dataset \\
arcsec & mag/arcsec$^2$ & mag/arcsec$^2$ &  $^a$       &         \\
\hline
\endfirsthead
\caption{continued} \\
\hline \hline
   r   & $\mu_V$        & $err_{\mu_V}$    & Technique & Dataset \\
arcsec & mag/arcsec$^2$ & mag/arcsec$^2$ & $^a$          &         \\
\hline
\endhead
\hline
\endfoot
    1.5 & 19.58  & 0.05  & AP  & ACS\\ 
    4.5 & 19.73  & 0.06  & AP  & ACS\\
    7.5 & 19.86  & 0.04  & AP  & ACS\\
   10.5 & 19.86  & 0.04  & AP  & ACS\\
   13.5 & 19.93  & 0.06  & AP  & ACS\\
   16.5 & 20.17  & 0.06  & AP  & ACS\\
   19.5 & 20.51  & 0.07  & AP  & ACS\\
   20.0 & 20.42  & 0.09  & SC  & ACS\\
   21.0 & 20.48  & 0.09  & SC  & ACS\\
   22.0 & 20.53  & 0.09  & SC  & ACS\\
   23.0 & 20.67  & 0.09  & SC  & ACS\\
   24.0 & 20.70  & 0.09  & SC  & ACS\\
   25.0 & 20.88  & 0.10  & SC  & ACS\\
   26.0 & 20.99  & 0.10  & SC  & ACS\\
   27.0 & 21.13  & 0.08  & SC  & ACS\\
   28.0 & 21.20  & 0.08  & SC  & ACS\\
   29.0 & 21.15  & 0.08  & SC  & ACS\\
   30.0 & 21.21  & 0.08  & SC  & ACS\\
   31.0 & 21.19  & 0.07  & SC  & ACS\\
   32.0 & 21.24  & 0.07  & SC  & ACS\\
   33.0 & 21.30  & 0.08  & SC  & ACS\\
   34.0 & 21.32  & 0.07  & SC  & ACS\\
   35.0 & 21.38  & 0.08  & SC  & ACS\\
   36.0 & 21.32  & 0.07  & SC  & ACS\\
   37.0 & 21.47  & 0.08  & SC  & ACS\\
   38.0 & 21.47  & 0.08  & SC  & ACS\\
   39.0 & 21.47  & 0.08  & SC  & ACS\\
   40.0 & 21.45  & 0.07  & SC  & ACS\\
   41.0 & 21.52  & 0.08  & SC  & ACS\\
   42.0 & 21.68  & 0.08  & SC  & ACS\\
   43.0 & 21.73  & 0.08  & SC  & ACS\\
   44.0 & 21.78  & 0.08  & SC  & ACS\\
   45.0 & 21.85  & 0.08  & SC  & ACS\\
   46.0 & 21.95  & 0.08  & SC  & ACS\\
   47.0 & 22.03  & 0.09  & SC  & ACS\\
   48.0 & 22.10  & 0.09  & SC  & ACS\\
   49.0 & 22.11  & 0.09  & SC  & ACS\\
   50.0 & 22.13  & 0.09  & SC  & ACS\\
   51.0 & 22.17  & 0.09  & SC  & ACS\\
   52.0 & 22.30  & 0.22  & SC  & S05\\
   54.0 & 22.29  & 0.22  & SC  & S05\\
   56.0 & 22.25  & 0.21  & SC  & S05\\
   58.0 & 22.33  & 0.21  & SC  & S05\\
   60.0 & 22.37  & 0.21  & SC  & S05\\
   62.0 & 22.40  & 0.21  & SC  & S05\\
   64.0 & 22.40  & 0.21  & SC  & S05\\
   66.0 & 22.47  & 0.22  & SC  & S05\\
   68.0 & 22.80  & 0.25  & SC  & S05\\
   70.0 & 23.21  & 0.29  & SC  & S05\\
   72.0 & 23.09  & 0.27  & SC  & S05\\
   74.0 & 23.06  & 0.26  & SC  & S05\\
   76.0 & 23.23  & 0.29  & SC  & S05\\
   78.0 & 23.42  & 0.31  & SC  & S05\\
   80.0 & 23.28  & 0.29  & SC  & S05\\
   82.0 & 23.31  & 0.28  & SC  & S05\\
   84.0 & 23.68  & 0.34  & SC  & S05\\
   86.0 & 24.05  & 0.38  & SC  & S05\\
   88.0 & 23.84  & 0.34  & SC  & S05\\
   90.0 & 23.87  & 0.35  & SC  & S05\\
   92.0 & 23.78  & 0.32  & SC  & S05\\
   94.0 & 23.70  & 0.33  & SC  & S05\\
   96.0 & 23.81  & 0.33  & SC  & S05\\
   98.0 & 23.96  & 0.35  & SC  & S05\\
  100.0 & 24.20  & 0.40  & SC  & S05\\
  102.0 & 24.24  & 0.41  & SC  & S05\\
  104.0 & 24.42  & 0.44  & SC  & S05\\
  106.0 & 24.42  & 0.44  & SC  & S05\\
  108.0 & 24.47  & 0.41  & SC  & S05\\
  110.0 & 24.47  & 0.41  & SC  & S05\\
  112.0 & 24.47  & 0.41  & SC  & S05\\
  114.0 & 24.37  & 0.42  & SC  & S05\\
  116.0 & 24.69  & 0.45  & SC  & S05\\
  118.0 & 24.57  & 0.45  & SC  & S05\\
  120.0 & 24.28  & 0.39  & SC  & S05\\
  122.0 & 24.20  & 0.36  & SC  & S05\\
  124.0 & 24.02  & 0.34  & SC  & S05\\
  126.0 & 23.78  & 0.30  & SC  & S05\\
  128.0 & 23.81  & 0.28  & SC  & S05\\
  130.0 & 23.99  & 0.33  & SC  & S05\\
  132.0 & 24.09  & 0.33  & SC  & S05\\
  134.0 & 24.20  & 0.36  & SC  & S05\\
  136.0 & 24.69  & 0.45  & SC  & S05\\
  138.0 & 25.35  & 0.60  & SC  & S05\\
  140.0 & 25.35  & 0.60  & SC  & S05\\
  142.0 & 25.35  & 0.60  & SC  & S05\\
  144.0 & 25.76  & 0.72  & SC  & S05\\
  147.5 & 25.25  & 0.11  & SC &  SDSS \\ 
  152.5 & 25.35  & 0.11  & SC &  SDSS \\ 
  157.5 & 25.46  & 0.12  & SC &  SDSS \\ 
  162.5 & 25.49  & 0.12  & SC &  SDSS \\ 
  167.5 & 25.55  & 0.12  & SC &  SDSS \\ 
  172.5 & 25.61  & 0.12  & SC &  SDSS \\ 
  177.5 & 25.69  & 0.12  & SC &  SDSS \\ 
  182.5 & 25.87  & 0.13  & SC &  SDSS \\ 
  187.5 & 25.91  & 0.13  & SC &  SDSS \\ 
  192.5 & 25.99  & 0.14  & SC &  SDSS \\ 
  197.5 & 26.02  & 0.14  & SC &  SDSS \\ 
  202.5 & 26.13  & 0.14  & SC &  SDSS \\ 
  207.5 & 26.27  & 0.15  & SC &  SDSS \\ 
  212.5 & 26.38  & 0.16  & SC &  SDSS \\ 
  217.5 & 26.59  & 0.17  & SC &  SDSS \\ 
  222.5 & 26.62  & 0.17  & SC &  SDSS \\ 
  227.5 & 26.75  & 0.18  & SC &  SDSS \\ 
  232.5 & 26.86  & 0.19  & SC &  SDSS \\ 
  237.5 & 27.12  & 0.21  & SC &  SDSS \\ 
  242.5 & 27.30  & 0.23  & SC &  SDSS \\ 
  247.5 & 27.57  & 0.27  & SC &  SDSS \\ 
  252.5 & 27.45  & 0.25  & SC &  SDSS \\ 
  257.5 & 27.57  & 0.26  & SC &  SDSS \\ 
  262.5 & 27.49  & 0.25  & SC &  SDSS \\ 
  267.5 & 27.57  & 0.25  & SC &  SDSS \\ 
  272.5 & 27.59  & 0.26  & SC &  SDSS \\ 
  277.5 & 27.62  & 0.26  & SC &  SDSS \\ 
  282.5 & 27.50  & 0.24  & SC &  SDSS \\ 
  287.5 & 27.43  & 0.23  & SC &  SDSS \\ 
  292.5 & 27.74  & 0.27  & SC &  SDSS \\ 
  297.5 & 27.66  & 0.25  & SC &  SDSS \\ 
  302.5 & 27.73  & 0.26  & SC &  SDSS \\ 
  307.5 & 27.92  & 0.29  & SC &  SDSS \\ 
  312.5 & 27.88  & 0.28  & SC &  SDSS \\ 
  317.5 & 27.90  & 0.28  & SC &  SDSS \\ 
  322.5 & 28.18  & 0.33  & SC &  SDSS \\ 
  327.5 & 28.20  & 0.33  & SC &  SDSS \\ 
  332.5 & 28.15  & 0.32  & SC &  SDSS \\ 
  337.5 & 28.50  & 0.39  & SC &  SDSS \\ 
  342.5 & 28.72  & 0.44  & SC &  SDSS \\ 
  347.5 & 28.64  & 0.42  & SC &  SDSS \\ 
  352.5 & 28.89  & 0.49  & SC &  SDSS \\ 
  357.5 & 29.06  & 0.54  & SC &  SDSS \\ 
  362.5 & 29.24  & 0.61  & SC &  SDSS \\ 
  367.5 & 29.11  & 0.55  & SC &  SDSS \\ 
  372.5 & 29.00  & 0.51  & SC &  SDSS \\ 
  377.5 & 29.02  & 0.51  & SC &  SDSS \\ 
  382.5 & 28.92  & 0.47  & SC &  SDSS \\ 
  387.5 & 29.07  & 0.52  & SC &  SDSS \\ 
  392.5 & 29.42  & 0.66  & SC &  SDSS \\ 
  397.5 & 29.12  & 0.53  & SC &  SDSS \\ 
  402.5 & 28.89  & 0.46  & SC &  SDSS \\ 
  407.5 & 29.17  & 0.54  & SC &  SDSS \\ 
  412.5 & 29.06  & 0.50  & SC &  SDSS \\ 
  417.5 & 28.96  & 0.47  & SC &  SDSS \\ 
  422.5 & 29.24  & 0.56  & SC &  SDSS \\ 
  427.5 & 29.27  & 0.57  & SC &  SDSS \\ 
  432.5 & 29.15  & 0.52  & SC &  SDSS \\ 
  437.5 & 29.67  & 0.75  & SC &  SDSS \\ 
  442.5 & 29.51  & 0.66  & SC &  SDSS \\ 
  447.5 & 29.53  & 0.67  & SC &  SDSS \\ 
  452.5 & 29.76  & 0.79  & SC &  SDSS \\ 
  457.5 & 30.04  & 0.97  & SC &  SDSS \\ 
  467.5 & 30.11  & 1.02  & SC &  SDSS \\ 
  472.5 & 30.15  & 1.04  & SC &  SDSS \\ 
  477.5 & 30.18  & 1.06  & SC &  SDSS \\ 
\hline	 
$^a$ AP  = Aperture Photometry; SC= Star Counts. \\
\end{longtable}
\end{scriptsize}
}
\end{document}